\documentclass[twocolumn, 10pt]{article}
\usepackage{physics}
\usepackage{amsmath}
\usepackage{amssymb}
\usepackage{bm}
\usepackage{siunitx}
\usepackage[a4paper, total={7in, 9in}]{geometry}
\usepackage{graphicx}
\usepackage{authblk}
\usepackage{abstract}
\usepackage{titlesec}
\usepackage[utf8]{inputenc}
\usepackage[T1]{fontenc}

\titleformat{\section}
  {\normalsize\bfseries}  
  {}       
  {0pt}              
  {}                 
\newcommand{\dis}{distinguishable }

\newcommand{\ind}{indistinguishable }

\title{Harnessing Photon Indistinguishability in Quantum Extreme Learning Machines}
\author[1]{Malo Joly}
\author[1,2]{Adrian Makowski}
\author[1, 6]{Baptiste Courme}
\author[3]{Lukas Porstendorfer}
\author[4]{Steffen Wilksen}
\author[5]{Edoardo Charbon}
\author[4]{Christopher Gies}
\author[6]{Hugo Defienne}
\author[1]{Sylvain Gigan}
\affil[1]{Laboratoire Kastler Brossel, École normale supérieure (ENS) – Université Paris Sciences \& Lettres (PSL), CNRS, Sorbonne Université, Collège de France, 24 rue Lhomond, Paris 75005, France}
\affil[2]{Institute of Experimental Physics, Faculty of Physics, University of Warsaw, Pasteura 5, 02-093 Warsaw, Poland}
\affil[3]{Institute for Theoretical Physics, University of Bremen, 28334 Bremen, Germany}
\affil[4]{Institute for Physics, Faculty V, Carl von Ossietzky University Oldenburg, 26129 Oldenburg, Germany}
\affil[5]{Advanced Quantum Architecture Laboratory (AQUA), Ecole Polytechnique Fédérale de Lausanne (EPFL)}
\affil[6]{Sorbonne Université, CNRS, Institut des NanoSciences de Paris, France}
\date{}
\begin{document}
\twocolumn[
\maketitle

\begin{abstract}
Recent advancements in machine learning have led to an exponential increase in computational demands, driving the need for innovative computing platforms. Quantum computing, with its Hilbert space scaling exponentially with the number of particles, emerges as a promising solution. In this work, we implement a quantum extreme machine learning (QELM) protocol leveraging indistinguishable photon pairs and multimode fiber as a random densely connected layer. We experimentally study QELM performance based on photon coincidences—for distinguishable and indistinguishable photons—on an image classification task. Simulations further show that increasing the number of photons reveals a clear quantum advantage. We relate this improved performance to the enhanced dimensionality and expressivity of the feature space, as indicated by the increased rank of the feature matrix in both experiment and simulation.
\end{abstract}
]

Advancements in machine learning have transformed various fields~\cite{lecun_deep_2015}. The exponential increase in computational demands reveals the limitations of classical computing for large-scale tasks which necessitate exploring alternative computing paradigms that offer enhanced capabilities~\cite{gasarch_alternative_2021}. Quantum computing offers a promising solution by leveraging quantum mechanics to perform computations intractable for classical computers~\cite{nielsen_quantum_2010}. Quantum systems exploit superposition and entanglement to process information in exponentially large Hilbert spaces, which provides inherent parallelism for accelerating machine learning algorithms and handling high-dimensional data.

Quantum Machine Learning (QML) models integrate quantum resources into machine learning frameworks~\cite{biamonte_quantum_2017}. Quantum Extreme Learning Machines (QELMs) extend classical Extreme Learning Machines (ELMs)\cite{huang_extreme_2006}—single-hidden-layer feed-forward neural networks with random hidden-layer weights and analytically determined output weights—by incorporating quantum systems~\cite{fujii_harnessing_2017}. They are particularly promising for the noisy intermediate scale quantum (NISQ) era as a complete control over the system is not needed.

Photonic systems are attractive for QELMs due to high-speed operation, low decoherence, and ease of manipulating photonic states~\cite{innocenti_potential_2023}. Previous photonic ELMs have used classical light and intensity measurements, showing potential for optical computing in machine learning~\cite{saade_random_2016, pierangeli_photonic_2021, lupo_photonic_2021}. Photon indistinguishability is a key quantum resource~\cite{aaronson_computational_2010}. Indistinguishable photons—identical in all degrees of freedom—can interfere quantum mechanically, leading to phenomena like the Hong-Ou-Mandel (HOM) effect~\cite{hong_measurement_1987}. This interference enables access to higher-dimensional Hilbert spaces, increasing computational model expressivity~\cite{innocenti_potential_2023, xiong_fundamental_2023, nerenberg_photon_2024}. While many theoretical studies of photonic QELMs exist, experimental demonstrations leveraging quantum resources for classical tasks are few, as most experiments focus on quantum tasks~\cite{suprano_experimental_2024, zia_quantum_2025}.

Here, we demonstrate a QELM by encoding data onto photon pairs, using a multimode fiber as a random mode mixer and measuring outputs using a single-photon avalanche diode (SPAD) camera, and investigate the impact of photon indistinguishability on QELM performance.

Our results show that coincidence-based ELMs outperform intensity-only ELMs. While the current experimental setup (2 photons and 22 detectors) does not reveal a clear advantage for indistinguishable compared to distinguishable photons, simulations show that indistinguishability significantly enhances QELM performance in Hilbert spaces at high dimensions, particularly with large number of photons. The improved performance arises from a favorable scaling of expressivity with dimensionality.

\begin{figure*}
    \centering
    \includegraphics[width=0.9\textwidth]{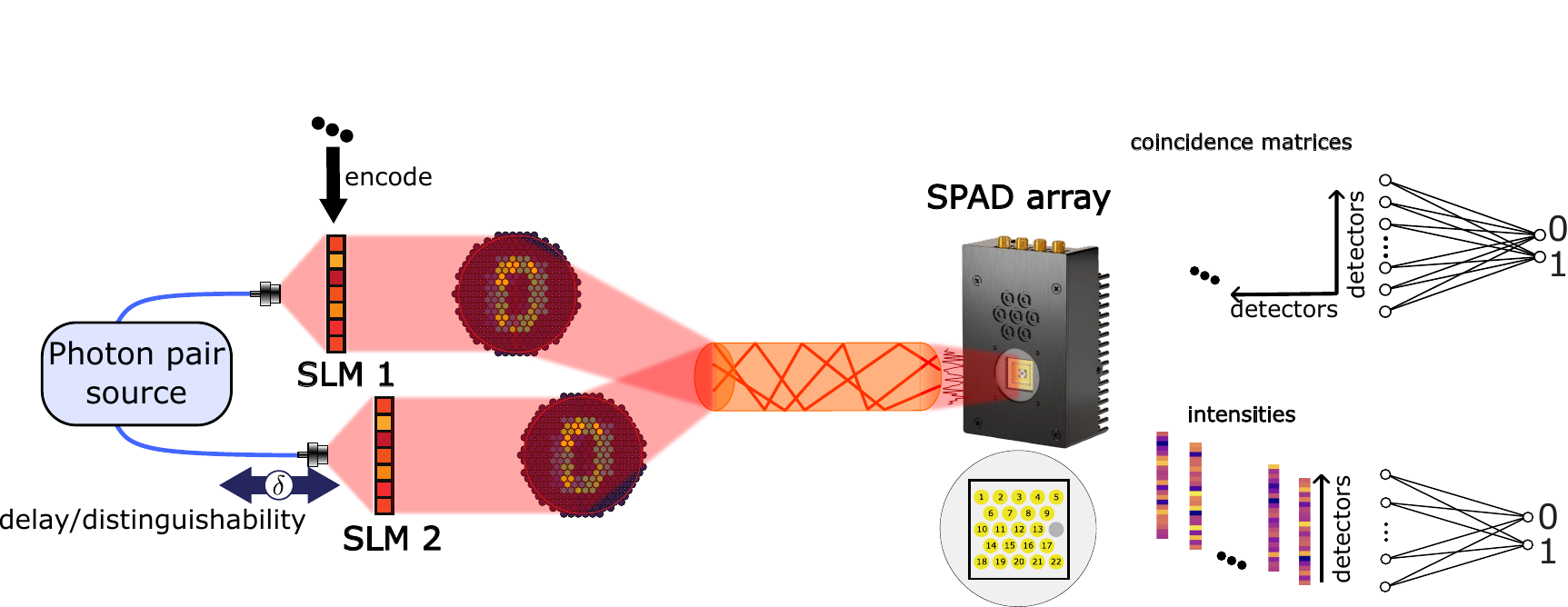}
    \caption{The setup is comprised of the quantum source and wavefront shaping setup. An electronic delay in one arm controls photon indistinguishability, measured using a fiber beam splitter and avalanche photodiode detectors, achieving 95\% HOM visibility after correcting for accidental coincidences. Photon pairs are injected into single-mode fibers and directed onto phase-only spatial light modulators (SLMs). After recombination at a polarizing beam splitter, photons are coupled into a \unit{50\micro\meter} multimode fiber (MMF) supporting 290 spatial modes. The input of the MMF is placed in the Fourier plane of the SLMs, enabling control of phase and amplitude for each spatial modes of both photons. The MMF output is measured by a time-stamping Single Photon Avalanche Detector (SPAD) array with 22 detectors, from which the coincidence matrix or intensity vector are reconstructed. The rate of accidental coincidence is of \unit{10^{-5}} per second and the crosstalk probability is of $0.1\,\%$.
    Our experiment implements an ELM by encoding input images on the SLM, the complex fixed random layer is provided by the MMF (with either intensity or coincidence detection) and a trainable electronic layer provides the classification. We encode 8-bit MNIST images, down-sampled to 8 by 8 pixels, onto the phase of each photon at the MMF input, interpolated to the MMF’s hexagonal input basis. We send 360 patterns of zeros and ones, acquiring timestamps for 2 minutes per pattern.
    Accuracies displayed are obtained by splitting our samples in training, confirmation and test sets in a $k$-fold cross validation, and with multiple permutation of the dataset.
    In the experiments, the error bars are over accuracies and obtained with $k$-fold permutation of the dataset in training-confirming and testing as well as sampling over subsets of detectors (100 at maximum). The error bars for the rank of feature matrices are obtained through sampling over subsets of detectors (maximum of 200).
    }
    \label{fig:1}
\end{figure*}

\emph{Extreme Learning Machines - }
Consider a non-linear random projection described by the mapping $x \mapsto P(x), \mathbb{R}^d \rightarrow \mathbb{R}^q$ with $d \ll q$—usually a matrix multiplication followed by a non-linear activation function. An ELM is a type of single hidden-layer feed-forward neural network where the input weights remain fixed (given by $P$), and the output weights are trained with a linear regression~\cite{huang_extreme_2006}.
An optical realization of an ELM can be realized by first encoding the data into the light, then letting it propagate in a complex medium, measuring the output intensity, and finally training a linear model~\cite{saade_random_2016}.
In the multiphoton version of this experiment that we considering in this paper, we replace the classical source by $n$-photon Fock states and training a linear model using coincidences measured between different output modes. The coincidences for $n$ photons and $m$ detectors writes as the sum of the contribution of distinguishable and indistinguishable photons,
\begin{equation}
\begin{aligned}
C_{j_1\cdots j_n} &= \alpha\frac{1}{n!} \abs{\sum_{\sigma \in S(n)} E_{\sigma(1) j_1} \cdots E_{\sigma(n) j_n}}^2\\
&+(1-\alpha)\frac{1}{n!} \sum_{\sigma \in S(n)} \abs{E_{\sigma(1) j_1} \cdots E_{\sigma(n) j_n}}^2,
\end{aligned}
\end{equation}
where $E_{ij}$ is the output field of the $i$-th photon at the $j$-th mode, $S(n)$ is the permutation group of $n$ elements, and $\alpha$ is the coefficient of indistinguishability ($\alpha=0$ for distinguishable photons and $\alpha=1$ for indistinguishable photons).
In this work, we call a \dis ELM (DELM), an ELM built with hidden layer given by coincidences of \dis photons, an \ind ELM (IELM), with \ind photons, an $\alpha$-partially distinguishable ELM ($\alpha$-ELM), with partially \dis photons of coefficient $\alpha$, and an intensity ELM (IntELM) built using only one photon (or equivalently measuring the intensity of multiple photons). Their respective random projections will follow the same terminology. For a task $\{x^{(k)}, y^{(k)}\}_{k=1}^p$, the \dis, \ind, $\alpha$ and intensity feature matrix, are obtained with the \dis, \ind, $\alpha-$, and intensity ELMs, respectively.

\begin{figure*}
    \centering
    \includegraphics[width=0.3\linewidth]{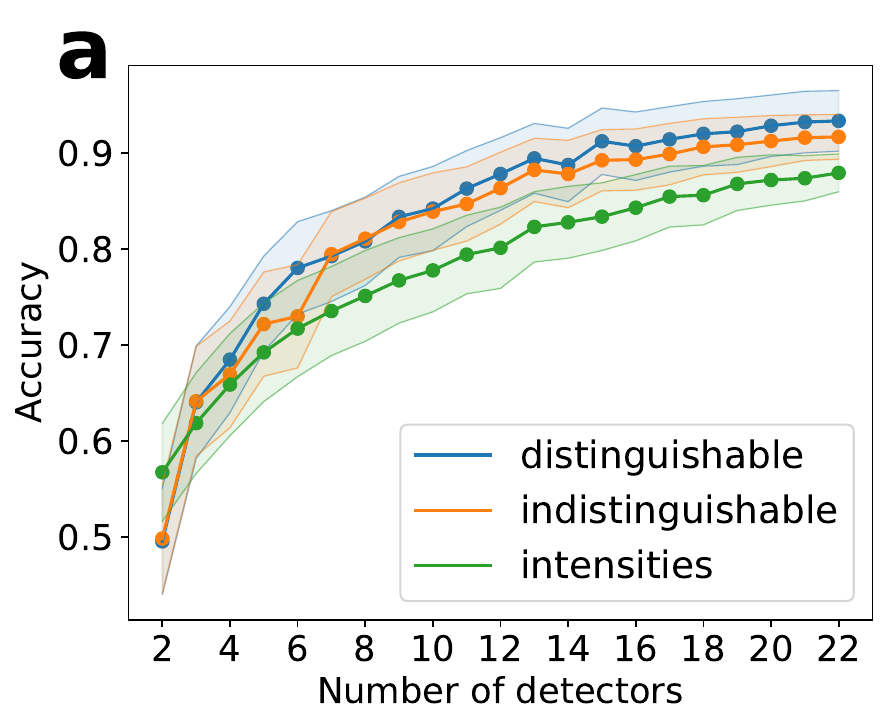}
    \includegraphics[width=0.3\linewidth]{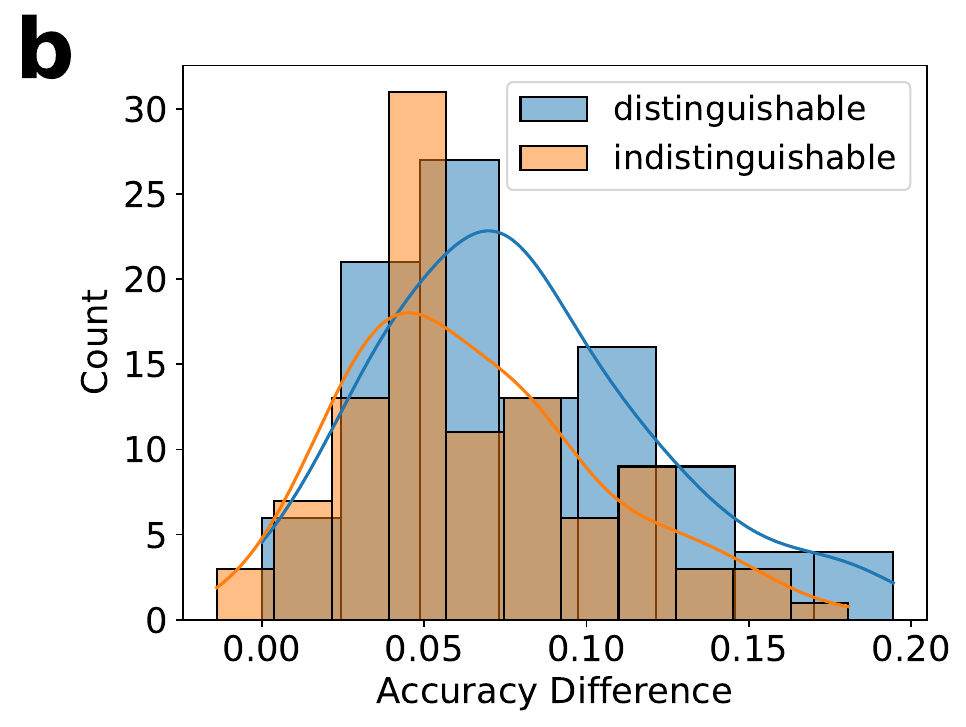}
    \includegraphics[width=0.3\linewidth]{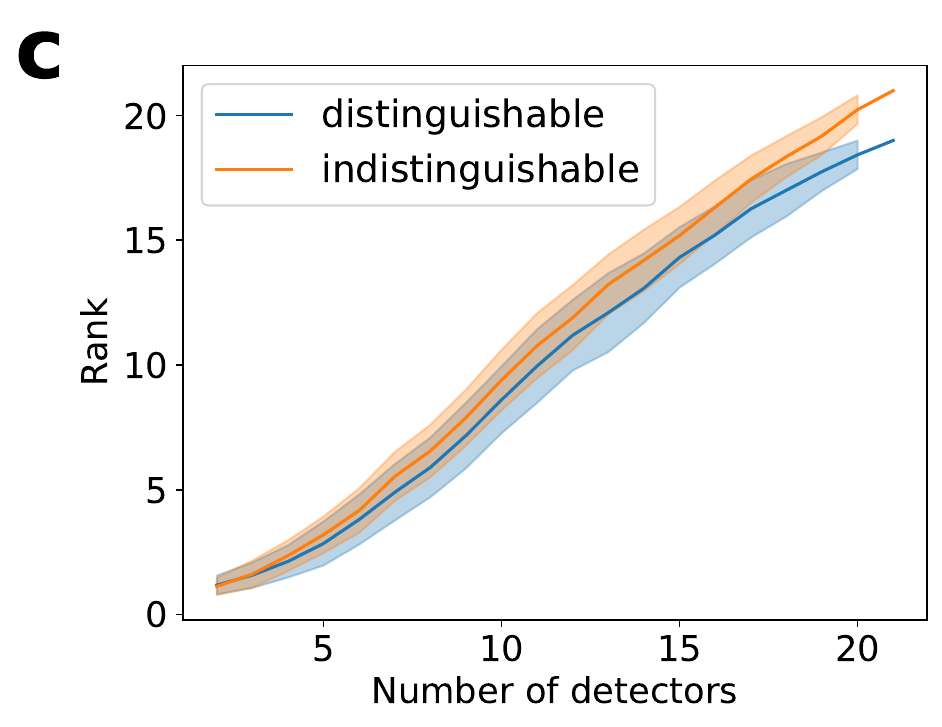}
    \caption{
    a: Scaling of the accuracy with the number of detectors used in the experimental QELM. Features built with \dis coincidences (blue), \ind coincidences (orange), intensities (green). The error-bar is the average over the the permutation of the dataset, and of sampling over subsets of detectors (for less than 22 detectors).
    b: Distribution of the accuracy difference for the \dis ELM (blue) and \ind ELM (orange). Each realization corresponds to a training-confirmation-testing for the coincidences features and intensities features of the same permutation.
    c: Scaling of the rank of the feature matrix with the number of detectors. Obtained when feeding the system with random input of dimension 300, and for 70 different random input. The rank is computed as the number of singular values above the 90\% threshold. Feature matrix obtained with the \dis photons (blue), \ind photons (orange). The error bar is over the sampling of subsets of detectors (for less detectors than 22).}
    \label{fig:2}
\end{figure*}

\emph{Experimental results with two photons and 22 detectors - }
We compare three experiments (Fig.~\ref{fig:1}), where we encode two classes, the zeros and ones, of the reduced MNIST data set on two photons~\cite{lecun_gradient-based_1998}, and measure the intensities, the coincidences of distinguishable photons or the coincidences of indistinguishable photons, with 2 photons and 22 detectors. In the three cases, the measurements are used as features of a linear classifier. Averaging over permutations of the dataset for 22 detectors, we achieve classification accuracies of 93\% with the DELM and 91\% with the IELM, significantly higher than the accuracy of IntELM of 87\% (see Fig.~\ref{fig:2}a), all well above the 50\% baseline of random guessing.
We observe a steady increase of the accuracy as we use more detectors for all three types of ELM: intensity-based, distinguishable photon-based, and indistinguishable photon-based (Fig.~\ref{fig:2}a). For five detectors and above, DELM and IELM outperform on average the intensity one.
The blue, orange and green error bars in Fig.~\ref{fig:2}, corresponding to the distinguishable photon, indistinguishable photon, and intensity-based ELMs, respectively, represent the standard deviation of the accuracy obtained when permuting the data set and subsets of detectors.

Our experimental results demonstrate that QELMs utilizing photon coincidences are capable of effectively solving classical machine learning tasks, such as the MNIST 0-1 classification task, here with a multimode fiber setup incorporating 22 detectors.

Moreover, it is possible to unambiguously compare the performance of coincidence-based and intensity-based ELMs. To do that, we fix for 22 detectors the permutations of dataset and train DELM, IELM, and IntELM. For a given permutation we compare the difference in accuracy for coincidence-based and intensity based ELMs (Fig.~\ref{fig:2}b) and find that measuring coincidences between photons, rather than using intensity measurements alone, consistently improves classification performance.

For the specific random transmission matrix used in the experiment, we observe higher classification performance for the DELM compared to the IELM (Fig.~\ref{fig:2}a). As our subsequent simulations demonstrate, this apparent advantage of DELM over IELM requires careful interpretation. The small dataset size combined with high variability across k-fold cross-validations results in standard deviations for all three approaches that are comparable to the observed differences between their performance curves.

The performance gain of coincidence-based ELMs is straightforward to understand, as the constructed ELM just has more neurons in its hidden layer. The DELM and IELM have the same number of modes: a difference in performance should come from the quality of their respective random projection. In line with literature ~\cite{hamid_compact_nodate, kar_random_2012, huh_low-rank_2023}, we propose to use the dimensionality of the random feature matrix as a measure of the system’s expressivity. The dimensionality can be captured by examining the number of singular values of the feature matrix above some threshold: consider a feature matrix $X \in \mathbb{R}^{p \times p}$—obtained with $\{x^{(k)}\}_{k=1}^p$ random inputs—with singular values $\sigma_i$. For a threshold $t \in [0, 1]$, the rank $r_t$ is defined as the smallest integer $r$ such that $t \leq \sum_{i=1}^r \sigma_i^2 \big/ \sum_{i=1}^p \sigma_i^2$, i.e., the number of singular values whose cumulative energy reaches at least a fraction $t$ of the total energy.
To generate such a feature matrix, we send 70 randomly generated data points of dimension equal to that of the multimode fiber (290 modes), recording the features derived from the measured coincidences of both the DELM and IELM.

We observe that the number of singular values (up to the 90\% threshold) is higher for the IELM than for the DELM as the number of detectors increases. The error bars indicate a statistically higher rank for the IELM above 22 detectors, along with an increasing rank gap. The dimensionality of the feature space, inferred from the rank of the feature matrix, shows a clear advantage for indistinguishable photons, even in the presence of experimental noise. This suggests that photon indistinguishability contributes to a richer feature space, which is likely to enhance learning in tasks requiring greater expressivity.

\begin{figure}
    \centering
    \includegraphics[width=0.8\linewidth]{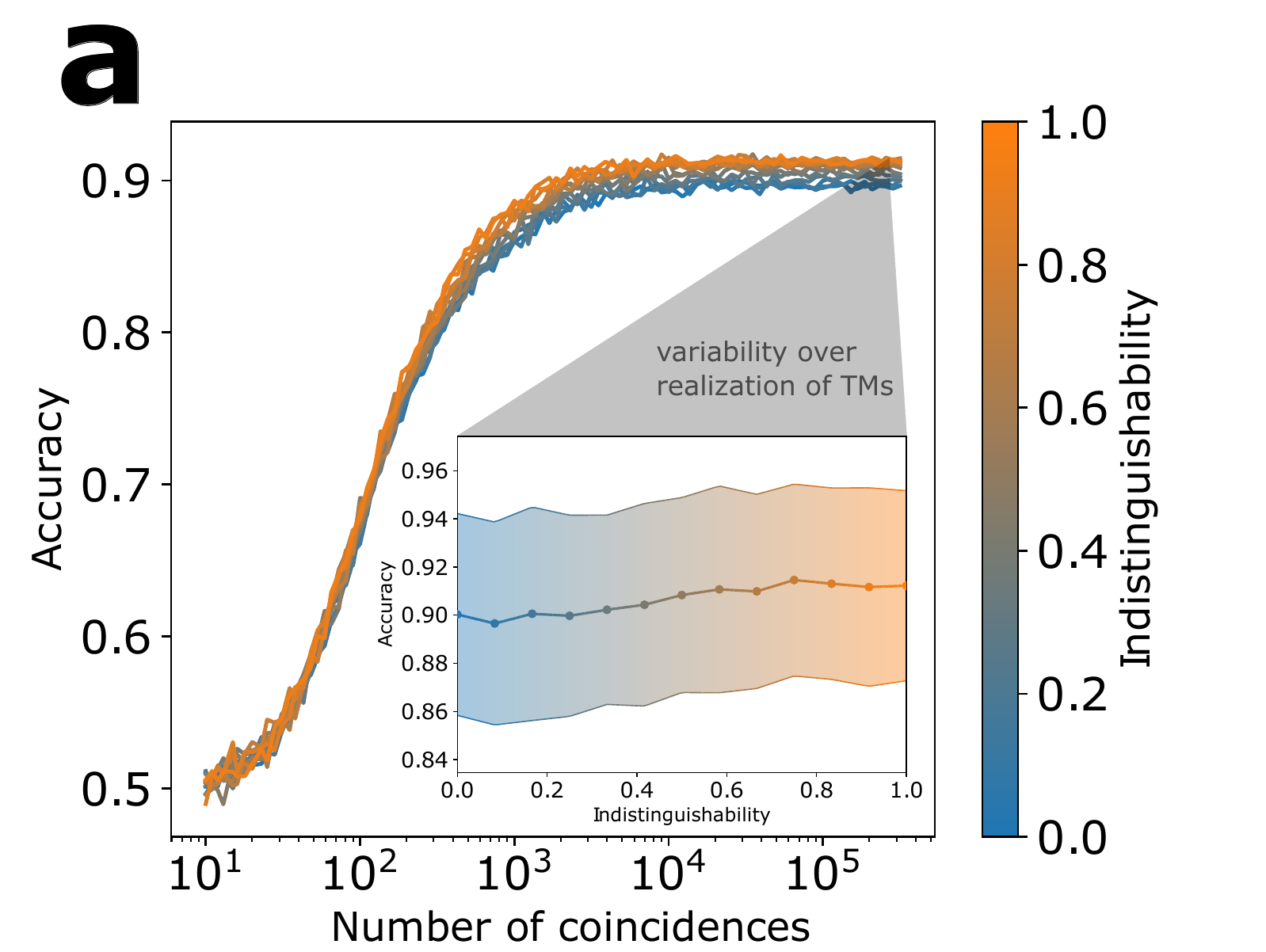}
    \includegraphics[width=0.8\linewidth]{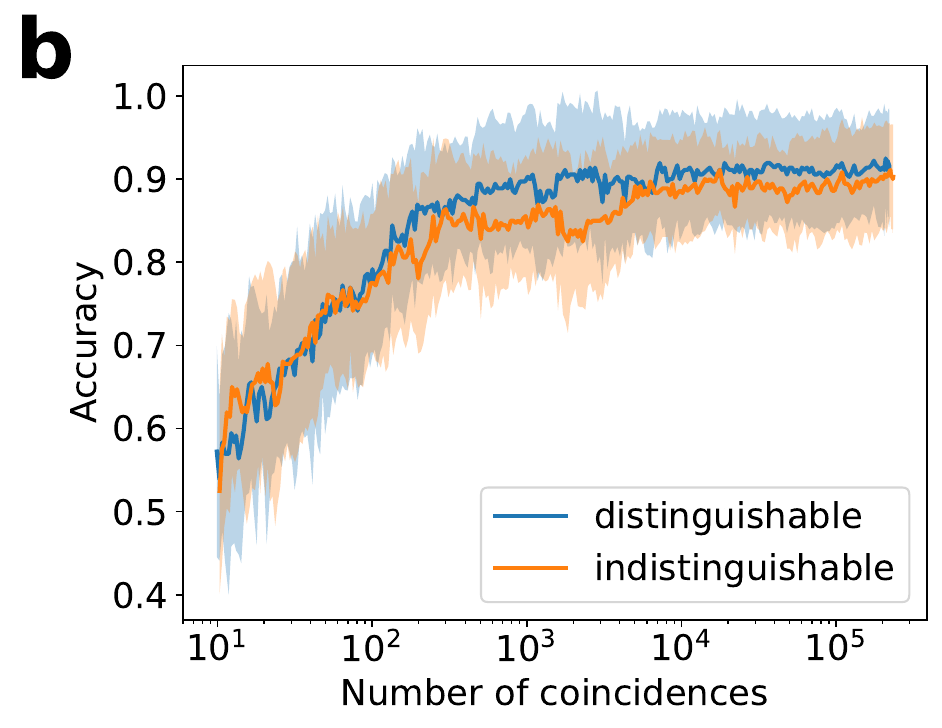}
    \caption{Simulation of the variability of the QELM and comparison with the experiment.
    a. accuracy vs coincidence count for different visibilities for the simulation of the experimental QELM with the MNIST 0-1 task and 360 data-points. Inset: accuracy vs. visibility for $10^5$ coincidence count. The error-bar is over the sampling of transmission matrices.\\
    b. Experimental accuracy vs. the number of coincidence count. For the \dis QELM (blue), \ind QELM (orange). The error-bar is over the permutation of the dataset.}
    \label{fig:3}
\end{figure}

\emph{Variability of the system - }
We simulate the experimental configuration and observe a steady increase in average performance as indistinguishability increases (until 0.73), indicating that indistinguishability continuously improves performance up to a certain point (Fig.~\ref{fig:3}a). However, for the dimension of the experiment, the standard deviation of accuracy due to the variability in the sampling of transmission matrices is far greater than the performance increase. At the experimental visibility of the indistinguishable photons (0.8), this variability prevents us from observing a statistical advantage of the IELM over the DELM, because the experiment is done for one transmission matrix only.
Performance variability is also evident in the experiment (Fig.~\ref{fig:3}b), especially when dealing with a small number of data points. We observe an increase in performance when reducing shot noise for both the DELM and IELM. In this particular realization of the random transmission matrix, the DELM outperforms the IELM; however, over permutations of the dataset, the standard deviations of the accuracies overlaps. 
Interestingly we find that an indistinguishability of 1 is not always optimal (shown here in the 0-1 task with an indistinguishability of 0.73). As the quality of the random projection dictates the performance of the ELM, we interpret this as the fact that partial indistinguishability results in a mix in the two random projections, augmenting the richness. This effect can be seen as an extension of works exploring the impact of partial coherence on photonic machine learning tasks~\cite{dong_partial_2024,pitruzzello_low-coherence_2024}.

\emph{Simulations when scaling the number of photons and modes - }
We study the scaling of the feature space and ELM performance with the number of photons and detectors, in a Hilbert space proposed in ~\cite{nerenberg_photon_2024}, for both distinguishable photons and indistinguishable photons.
We simulate the ELM while increasing the number of photons from one to five, using a fixed number of detectors ($m=16$), for a five-class classification task on the FashionMNIST dataset~\cite{xiao_fashion-mnist_2017} and show the classification accuracy with respect to the number of photons. (Fig.~\ref{fig:3}a). For the IELM, we observe a steady increase in performance from 60\% to 70\% as the number of photons increases. For the DELM, performance increases from 60\% to 64\% when increasing the number of photons from one to two, but then drops slightly to 63\% when increasing to five photons. From three photons onward, the standard deviation for the IELM and DELM separates, indicating a statistical significance of indistinguishable photons over distinguishable ones.

We now turn from classification accuracy to studying the rank of the feature matrix to understand more in depth the gap in performance. In simulation, we renormalize the value of the rank computed at threshold $t$ with the rank of a Gaussian random matrix of size $p$, computed with the same threshold $t$, giving the value of the full rank.

\begin{figure*}
    \includegraphics[width=0.33\linewidth]{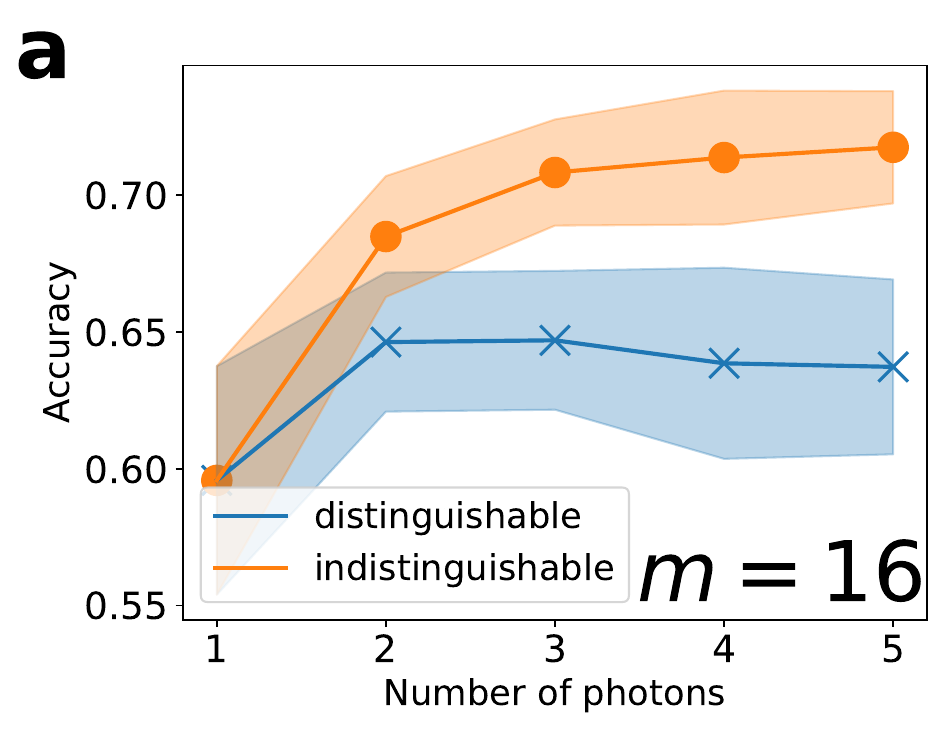}\label{fig:4a}
    \includegraphics[width=0.33\linewidth]{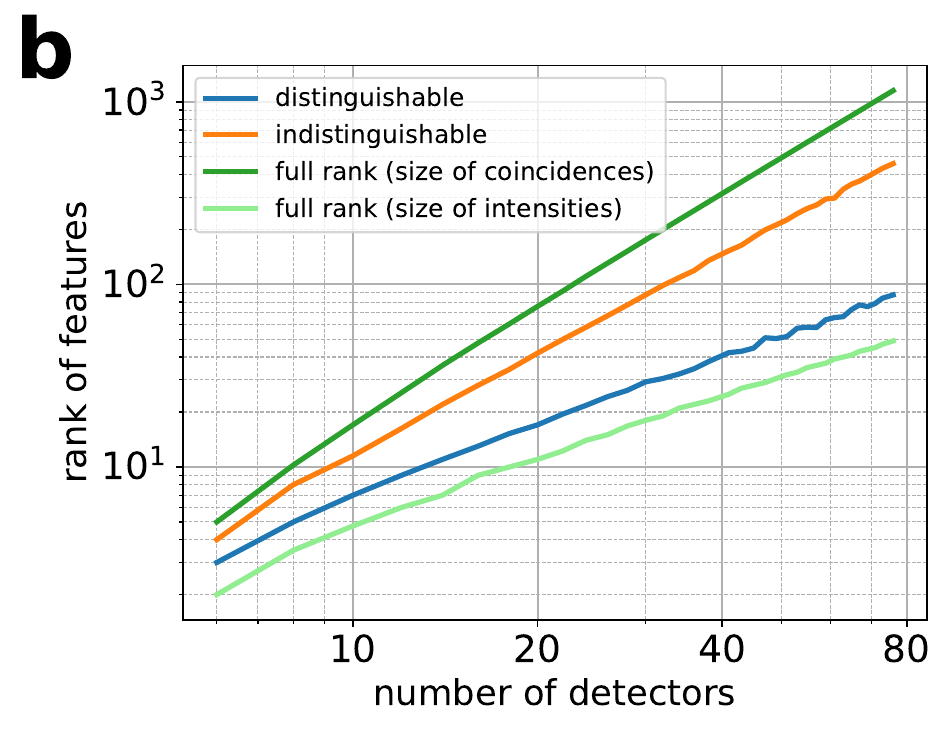}\label{fig:4b}
    \includegraphics[width=0.33\linewidth]{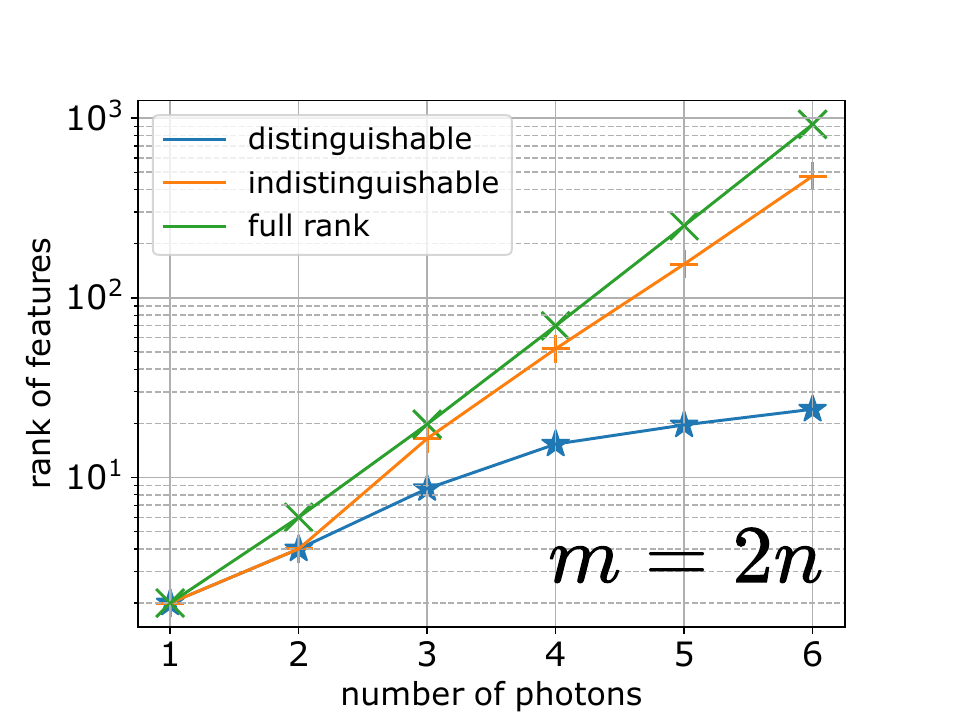}\label{fig:4c}
    \caption{Scaling of the performance with the dimensionality.\\
    a. accuracy scaling with the number of photons for the 1-5 FashionMNIST task with 4000 samples for \dis QELM (blue) and \ind QELM (orange). We fix the number of detectors to be $m=16$. The error-bar is over the sampling of TMs.\\
    b. scaling of the rank of the system with the number of detectors (log-log scale), for two photons. The rank is measured as  the rank of the feature matrix when feeding the system with random input, \dis in blue, the \ind in orange. We plot the dimension of the coincidence feature space (dark green) and intensity feature space (light green).\\
    c. scaling of the rank of the system with the number of photons (ordinate-log scale), for a number of photons $n$, the number of detectors is chosen to be $m=2n$. The rank is measured as  the rank of the feature matrix when feeding the system with random input. The \dis in blue, the \ind in orange. We plot the dimension of the $n$-fold coincidence space in green.}
    \label{fig:4}
\end{figure*}

In Fig.~\ref{fig:4}b, we present on a log-log scale the functions $m(m-1)/2$ (dark green) and $m$ (light green), alongside the rank of the feature matrices, generated by random inputs, for both distinguishable and indistinguishable photons. We observe that the rank of the indistinguishable feature matrix exceeds that of the distinguishable one and exhibits a different scaling behavior, as indicated by their respective slopes in the log-log plot. Specifically, while the rank of the distinguishable feature matrix scales linearly with the number of detectors $m$, the rank of the indistinguishable feature matrix scales in between linear and quadratic.
While the feature space is still richer even in the presence of noise, our measurement of dimensionality—based on sending random data through the system— is impacted by the limited coincidence count. As shown in the supplementary figures, a high amount of shot noise in the system will add up in the measured feature matrix and the its rank will converge to the rank of a random matrix-rank equal to the dimension of the matrix—for high shot noise (low coincidence count). This noise may explain the difference in scaling between theory (Fig.~\ref{fig:4}b) and experiment (Fig.~\ref{fig:2}c).

This distinct scaling behavior is also evident when we increase the number of photons. The dimensionality of the $n$-fold coincidence space for $m$ detectors is given by the binomial coefficient $\binom{m}{n}$. This time we scale the number of detectors with the number of photons $m=2n$, as the maximal binomial coefficient scales as $\binom{2n}{n} \sim 2^n$ through Sterling's formula. With the ordinate in log scale, we find an exponential scaling of the rank of features for the indistinguishable ELM, and a polynomial scaling for the distinguishable case.

The rank of the feature matrix serves as a practical proxy of system expressivity and demonstrates a clear scaling advantage for indistinguishable photons as the number of detectors and photons increases. We can link the exponential increase of the Hilbert space-inherent to quantum systems-as an exponential increase in the expressivity of the ELM it underpins. Notably, this measure of expressivity remains robust against the variability introduced by different TM realizations, providing a reliable indicator of the system’s ability to extract information from random inputs. Our findings show that even with two photons, IELMs exhibit better scaling behavior compared to DELMs, with a particularly clear advantage when increasing the number of photons over increasing the number of detectors (supplementary Fig.~\ref{fig:7}). This improvement is directly linked to the richness of quantum interference phenomena, which enhances the dimensionality of the system’s Hilbert space. 

Figure~\ref{fig:4}a shows an increase in performance from one photon to two distinguishable photons, with no clear sign of dimensionality increase. This performance gain can be attributed to an increase in nonlinearity: Indeed, it has has been demonstrated that a fixed weights encoder with higher non-linearity exhibits better performance \cite{xia_nonlinear_2024}. Viewing our optical system as a non-linear encoder, the one photon encoder's non-linearity is point-wise: the encoding function and measurement process. The two photon encoder has an extra non-linearity coming from the tensor product of the data.

It should be emphasized that our current approach requires the measurement of the entire Hilbert space, which serves as activation of the hidden layer of the ELM, while practical in understanding the different behavior of the ELM with indistinguishability, this approach is not scalable to a high number of photons where the size of the Hilbert space grows exponentially. Nevertheless, theoretical studies of QELM, using POVM of the Hilbert space, suggest that the gain in expressivity is still present, due to the complex nature of the quantum interference - even when restricting to a few coincidence modes~\cite{xiong_fundamental_2023}.

In conclusion, we have demonstrated that photon indistinguishability can be leveraged as a quantum resource to enhance the performance of quantum extreme learning machines. By constructing ELMs based on photon coincidences rather than solely on intensity measurements, we showed experimentally that these quantum-enhanced models improve the accuracy of machine learning tasks. While our experimental results with two photons and a limited number of detectors did not reveal a significant difference between distinguishable and indistinguishable photons in low-dimensional systems, simulations indicated that increasing the dimensionality—particularly by adding more photons—leads to a clear advantage for indistinguishable photons. This advantage arises from the enhanced expressivity and dimensionality of the feature space, as evidenced by the increased rank of the feature matrix.
Our findings highlight the potential of photon indistinguishability to enrich the feature space of quantum machine learning models without relying on high-dimensional entanglement. The scalability of QELMs using indistinguishable photons suggests promising avenues for future research, particularly in developing more efficient quantum machine learning algorithms capable of handling higher-dimensional data and more complex tasks.
However, our current approach requires measuring the entire Hilbert space, which is impractical as the number of photons increases due to exponential scaling. Future work could focus on strategies to harness the benefits of photon indistinguishability without requiring full Hilbert space measurement, such as utilizing partial measurements or exploiting specific quantum properties to maintain expressivity with fewer resources.

\section{Methods}
\emph{Experimental apparatus - } Supplementary figure~\ref{fig:5} shows our experimental setup, comprising the quantum source (a1) and wavefront shaping setup (a2). A 405,nm continuous-wave laser pumps a ppKTP crystal, generating photon pairs via type II SPDC in two Gaussian modes. An electronic delay in one arm controls photon indistinguishability, measured using a fiber beam splitter and avalanche photodiode detectors, achieving 95\% HOM visibility after correcting for accidental coincidences. Photon pairs are injected into single-mode fibers and directed onto phase-only spatial light modulators (SLMs). After recombination at a polarizing beam splitter, photons are coupled into a \unit{50\micro\meter} multimode fiber (MMF) supporting 290 spatial modes. The input of the MMF is placed in the Fourier plane of the SLMs, enabling control of phase and amplitude for each spatial modes of both photons. The MMF output is measured by a time-stamping SPAD array with 22 detectors, from which the coincidence matrix is reconstructed~\cite{leedumrongwatthanakun_programmable_2020, makowski_large_2023}.

\emph{Experimental task - }
Our experiment implement an ELM by encoding input images on the SLM, the complex fixed random layer is provided by the MMF (with either intensity or coincidence detection) and a trainable electronic layer provides the classification. We encode 8-bit MNIST images, down-sampled to 8 by 8 pixels, onto the phase of each photon at the MMF input, interpolated to the MMF’s hexagonal input basis (Fig.~\ref{fig:1}). We send 360 patterns of zeros and ones, acquiring timestamps for 2 minutes per pattern.
Accuracies displayed are obtained by splitting our samples in training, confirmation and test sets in a $k$-cross validation, and with multiple permutation of the dataset. We simulate this task in Fig.~\ref{fig:3}.
In the experiment of Fig.~\ref{fig:2}a, the error bars are obtained with $k$-fold permutation of the dataset in training-confirming and testing as well as sampling over subsets of detectors (100 at maximum for processing time reasons). In the experiment of Fig.~\ref{fig:2}c, the rank of the feature matrix is independent of the permutation of the dataset, and the error bars are obtained through sampling over subsets of detectors (maximum of 200). For 22 number of detectors, there is only one subset of detectors, and thus we do not know the error bar.
In simulations, they are obtained through sampling over many (100 in all simulations) random TMs and simulating an experiment for each one.\\
\emph{Scaling simulations task - }
Simulations of Fig.~\ref{fig:4} are for classifying 5 classes of the FashionMNIST dataset (with various number of samples).

\section*{Acknowledgement}
This study has been financed by the Association National de la Recherche and the Deutsche Forschungsgemeinschaft (German
Science Foundation, DFG) via the project PhotonicQRC as well as the Swiss National Foundation, Sinergia grant CRSII5\_216600 LION: Large Intelligent Optical Networks.

\begingroup
\footnotesize  
\bibliographystyle{unsrt}
\bibliography{biblio}
\endgroup

\end{document}